\documentclass[12pt,intlim,righttag]{article}
\usepackage{amsmath,amsthm}
\usepackage{amssymb}
\usepackage{amsfonts}

\newcommand{\essinf}{\operatornamewithlimits{ess\,inf}}
\newcommand{\esssup}{\operatorname{ess\,sup}}

\newcommand{\la}{\langle}
\newcommand{\ra}{\rangle}
\newcommand{\wt}{\widetilde}

\newcommand{\sg}{\sigma}
\newcommand{\ld}{\lambda}

\newcommand{\vps}{\varepsilon}

\newcommand{\ift}{\infty}
\newcommand{\be}{\begin}
\newcommand{\ee}{\end}
\newcommand{\lbl}{\label}

\newcommand\beq{\begin{equation}}
\newcommand\eeq{\end{equation}}

\theoremstyle{Theorem}

\theoremstyle{corollary}

\theoremstyle{remark}

\newtheorem{assumption}{Asumption}

\theoremstyle{definition}

\begin{document}
\title{On regularity of primal and dual dynamic value functions related to  investment problem
 }

\author{M. Mania and R. Tevzadze}

\date{~}
\maketitle

\begin{abstract}
{\bf Abstract.} We study regularity properties of the dynamic value functions  of primal and dual problems of optimal investing  for utility functions defined on the whole
real line.  Relations between decomposition terms of value processes of  primal and dual problems and  between optimal solutions of basic and conditional  utility maximization problems are established.  These properties are used to show that the value function satisfies a corresponding  backward stochastic partial differential equation.
 In the case of complete markets we give conditions on the utility function when this equation admits a solution.

\end{abstract}

\bigskip


\


\section{Introduction}

We consider a financial market model, where the dynamics of asset prices is described by the continuous semimartingale $S$  defined on the
complete  probability space $(\Omega,\mathcal{F}, P)$  with  continuous filtration $F=({F}_t,t\in[0,T])$, where $\mathcal{F}=F_T$ and $T<\infty$.
We work with discounted terms, i.e. the bond is assumed to be a constant.

Denote by ${\mathcal M}^e$ (resp. ${\mathcal M}^a$)
the set of probability  measures $Q$ equivalent (resp. absolutely continuous with respect) to $P$
such that $S$ is a local martingale under $Q$.

Throughout the paper we assume that the  filtration $F$ is continuous (i.e. all $F$-local martingales are continuous) and
\begin{equation}\label{mm}
{\mathcal M}^e\neq\emptyset.
\end{equation}
The continuity of $F$ and the existence of an equivalent martingale measure imply that the structure condition is satisfied, i.e. $S$ admits the decomposition
$$  S_t=M_t+\int_0^t\lambda_s \,d\langle M\rangle_s,\quad  \int_0^t \lambda^2_s\,d\langle M\rangle_s <\infty      $$
for all $t$ $P$-a.s., where $M$ is a continuous local martingale and $\lambda$ is a predictable process.

 Let $U=U(x):R\to R$ be a utility function taking finite values at all points of real line $R$ such that $U$ is continuously
differentiable, increasing, strictly concave and satisfies the Inada conditions
\begin{equation}\label{inada}
    U'(\infty)=\lim_{x\to\infty}U'(x)=0,\quad U'(-\infty)=\lim_{x\to -\infty}U'(x)=\infty.
\end{equation}
We also assume that $U$ satisfies the condition of reasonable  asymptotic elasticity (see \cite{KSch} and \cite{S3}), i.e.
\begin{equation}\label{ae}
    \limsup_{x\to\infty}\frac{xU'(x)}{U(x)}<1,\quad \liminf_{x\to -\infty}\frac{xU'(x)}{U(x)}>1.
\end{equation}

We consider the utility maximization problem, i.e. the problem of finding a trading strategy $(\pi_t,t\in[0,T])$ such that the expected utility of terminal wealth $X_T^{x,\pi}$ becomes maximal. The wealth process, determined by a self-financing
trading strategy $\pi$ and initial capital $x$,  is defined as a stochastic integral
$$
X^{x,\pi}_t=x+\int_0^t\pi_udS_u,\;\;\;0\le t\le T.
$$
The predictable, $S-$integrable  process $\pi$  we call admissible
if the stochastic integral $(\int_0^t\pi_udS_u, t\in[0,T])$ is uniformly bounded
from below.

The value function $V$ associated to the problem is given by
\begin{equation}\label{ut1}
    V(x)=\sup_{\pi\in\Pi} E\bigg[ U\bigg(x+\int_0^T\pi_u\,dS_u\bigg)\bigg],
 \end{equation}
where $\Pi$ is the class of admissible strategies.

For  the utility function $U$ we denote
by $\wt U$ its  convex conjugate
\begin{equation}\label{dual}
\wt U(y)=\sup_x(U(x)-xy),\;\;\;\;y>0.
\end{equation}
The dual problem to (\ref{ut1}) is
\begin{equation}\label{dualv}
\wt V(y)=\inf_{Q\in{\cal M}^e}E[\wt U(y\rho_T^Q)],\;\;\;y>0,
\end{equation}
where $\rho^Q_t=dQ_t/dP_t$ is the density process of the measure $Q\in{\cal M}^e$ relative to the basic measure $P$.

Let $\tau$ be a stopping time valued in $[0,T]$.
Denote by $\Pi_\tau$ the class of admissible processes, such
that $\pi=\pi 1_{[\tau,T]}$.
Define ${\cal Z}_{\tau,y}=\{Y:Y=y\frac{\rho_T}{\rho_\tau},\; \rho_T=\frac{dQ}{dP},\; Q\in{\cal M}^e(S)\}$.

The dynamic value functions of primal and dual problems are
 defined as
\begin{equation}\label{ut5}
    V(\tau,x)=\esssup_{\pi\in\Pi_\tau} E\bigg[U\bigg(x+\int_\tau^T\pi_u\,dS_u\bigg)\bigg|{F}_t\bigg],
\end{equation}
\begin{equation}\label{utd}{\widetilde V}(\tau,y)=\essinf_{Y\in {\cal Z}_{\tau,y}}E\Big[{\widetilde U}(Y)\mid F_t\Big],\;\;\;y>0.
\end{equation}
 For $V(0,x)$ and $\wt V(0,y)$ we use the notations $V(x)$ and $\wt V(y)$ respectively.
Following \cite{S3} we make

\begin{assumption}\label{as1}
 For each $y>0$ the dual value function ${\widetilde V(y)}$
is finite and the minimizer $Q^*(y)\in {\mathcal{M}}^e$ (called the minimax martingale measure) exists.
\end{assumption}
Let $\Pi_x$ be the class  of predictable $S$ integrable processes
$\pi$ such that $U(x+(\pi\cdot S)_T)\in L^1(P)$ and $\pi\cdot S$ is
a supermartingale
under each $Q\in {\cal M}^a$ with finite $\wt U$-expectation
$E{\wt U}(\frac{dQ}{dP})$, where
the notation $\pi\cdot S$ stands for the stochastic integral.

Denote $Q(x)=Q^*(y)=Q^*(V'(x))$.

It was proved in \cite{S1} that  optimal strategy
$\pi(x)\in\Pi_x$ of problem
 (\ref{ut1}) exists, is unique and  $V(x)=EU(X_T(x))$,
where the optimal wealth $X_T(x)=x+\int_0^T\pi_u(x)\,dS_u$
is a uniformly integrable
$Q(x)$-martingale.

Besides, the following duality
 relations hold true almost surely
\begin{equation}\label{main}
    U^{'}(X_T(x))=Z_T(y), \;\; y=V^{'}(x),
\end{equation}
\begin{equation}\label{main2}
    V'\bigg(t, x+\int_0^t\pi_u(x)\,dS_u\bigg)=Z_t(y),\quad t\in [0,T],
\end{equation}
where $y=V'(x)$ ( (see \cite{S3} and  Proposition A3 from \cite{MT15} for the dynamic version). Hereafter we shall use these results without further comments.

Our goal is to study the properties of the dynamic value function $V(t,x)$ and the optimal wealth process $X_t(x)$.
It is well known (see e.g., \cite{MT10}) that for
any $x\in R$ the process $(V(t,x),t\in [0,T])$ is a supermartingale admitting an RCLL (right-continuous with left limits) modification.

Therefore, using the Galchouk--Kunita--Watanabe (GKW) decomposition, the value function is represented as
\be{equation}\label{value}
 V(t,x)=V(0, x)- A(t,x)+\int_0^t\psi(s,x)\,dM_s+L(t,x),
\ee{equation}
where for any $x\in R$ the process $A(t,x)$ is  increasing  and $L(t,x)$ is a local martingale orthogonal to $M$.

Let consider the following assumptions:

\begin{itemize}
\item[a)] $V(t,x)$ is  two-times continuously differentiable at $x$  $P$- a.s. for any $t\in [0,T]$,

\item[b)] for any $x\in R$  the process $V(t,x)$ is a special semimartingale with bounded variation part absolutely continuous with respect to $\langle M\rangle$, i.e.
$$  A(t,x)=\int_0^ta(s,x)\,d\langle M\rangle_s,       $$
for some real-valued function $a(s,x)$ which is predictable  and $\langle M\rangle$-integrable for any $x\in R$,

\item[c)] for any $x\in R$ the process $V'(t,x)$ is a special semimartingale with the decomposition
$$  V'(t,x)=V'(0, x)- \int_0^ta'(s,x)\,d\langle M\rangle_s+\int_0^t\psi'(s,x)\,dM_s+L'(t,x),     $$
where $V'$, $a'$, $\psi'$ and $L'$ are partial derivatives at $x$ of $V$, $a$, $\psi$ and $L$, respectively.
\end{itemize}

We shall say that $(V(t,x),t\in[0,T])$ is a regular family of semimartingales if for $V$ conditions a), b) and c) are satisfied.

We  shall consider  also the conditions:

\begin{itemize}
\item[d)] the conditional optimization problem (\ref{ut5}) admits a solution, i.e., for any $t\in[0,T]$ and $x\in R$ there exists a strategy $\pi(t,x)$ such that
\begin{equation}\label{cond}
V(t,x)=E\big(U(x+\int_t^T\pi_u(t,x)dS_u)|F_t),
\end{equation}

\item[e)] for each $s\in [t,T]$  the function $(X_s(t,x)=x+\int_t^s\pi_u(t,x)dS_u,\;s\ge t)$ is  continuous
at $(t,x)$ $P-$a.s. .
\end{itemize}

It was shown in \cite{MT3, MT8, MT10} that if the value function  satisfies conditions a)-e) then it solves
the following backward stochastic partial differential equation (BSPDE)
\begin{align}
    V(t,x)= V(0,x) & +\frac{1}{2}\int_0^t \frac{(\varphi'(s,x)+\lambda(s)V'(s,x))^2} {V''(s,x)}\,d\langle M\rangle_s \notag \\
    & +\int_0^t\varphi(s,x)\,dM_s+L(t,x),\quad  V(T,x)=U(x). \label{ut6}
\end{align}

Our aim is to study conditions on the basic objects (on the asset price model and on the objective function $U$) which will guaranty that
the value function $V(t,x)$ is a regular family of semimartingales and conditions d) and e) are also satisfied, in order to show that the solution of equation (\ref{ut6}) exists. In Theorem 3 of section 5 we provide such type conditions in the case of complete markets.

The main example, where all conditions a)-e) are satisfied
is the case of exponential utility function
$
U(x)=-e^{-\gamma x}
$
with risk aversion parameter $\gamma\in(0,\infty)$. In this case the corresponding  value function
is of the form $V(t,x)=-e^{-\gamma x}V_t$, where $V_t$
is a special semimartingale. Besides,  $\widetilde{U}(y)=\frac{y}{\gamma}\big(\ln\frac{y}{\gamma}-1\big)$ and Assumption 1  is equivalent to the existence of $Q\in\mathcal{M}^e$ with finite relative entropy $EZ^Q_T\ln Z^Q_T$ (see e.g.  \cite{BF}).

We first investigate whether Assumption 1 implies an existence
of an optimal strategy to the
conditional maximization problem (\ref{ut5}) and how is  this
strategy  related
to the optimal strategy of the basic problem (\ref{ut1}).

It was shown in \cite{S3} that if we start at time $\tau$ with the optimal wealth
$X_\tau(x)$ then the optimal value in
(\ref{ut5}) is attained by
$\pi(\tau, x)=\pi(0,x)I_{]\tau, T]}$,
i.e.,
$$
E[U(X_T(x))|F_\tau]\ge E[U(X_\tau(x)+\int_\tau^T\pi_udS_u)
\mid F_\tau],\;\pi\in\Pi_{\tau},
$$
which is well understood from the Bellman Principle.

Under additional conditions we shall show (see Theorem 1) that if we start at time $\tau$ with the wealth
equal to arbitrary amount $x$, then the optimal strategy
$\pi(\tau, x)$ of (\ref{ut5}) is expressed in terms of the
optimal strategy
$\pi(x)=\pi(0,x)$ and the optimal wealth $X_\tau(x)= X_\tau(0,x)$
of (\ref{ut1}) at time $\tau$ by the equality
$$
\pi_t(\tau, x)=\pi_t(X_\tau^{-1}(x)), \;\;\;t\ge\tau\;\;\;\mu^{\la S\ra}-a.e.,
$$
where $X^{-1}_t(x)$ is the inverse of the optimal wealth $X_t(x)$.


In section 3 we establish relation between decomposition terms of the value process $V(t,x)$ (\ref{value}) with corresponding terms of the dual value process $\widetilde V(t,y)$.

The problem related with condition a) was studied in \cite{KS} for utility functions defined on the positive real line
for value functions at time $0$ and  in \cite{MT15} for dynamic value function $V(t,x)$ corresponding to utility functions defined on the whole real line.

The problems related with conditions b) and c) we connect with an existence of the inverse flow $X_t^{-1}(x)$ of the optimal
wealth. In \cite{MT15} conditions are given when for any $t$  the optimal wealth is an increasing function of $x$ $P$-a.s.
and that an adapted inverse of $X_t(x)$ exists. In Proposition 2 of section 4 we derive a stochastic differential equation for  the inverse of the optimal wealth $\psi_t(x)=X_t^{-1}(x)$ and based on this result  we give in  Proposition 3  sufficient conditions when  b) and c) are fulfilled.

Finally in Section 5 in the case of complete markets we give conditions on utility function for which all conditions a)-e) are fulfilled and
 the value function  $V(t,x)$ satisfies $BSPDE$ (\ref{ut6}).

In the paper \cite{hor} a new approach  was developed, where the solution of the problem (\ref{ut1}) was reduced
to the solvability of a system of Forward-Backward equations which is also a heavy task. Note that they showed that in case of complete markets this system admits a solution
under conditions similar to condition r1) of section 5.

\section{The relation between the basic and conditional utility
maximization problems}

In this section we study basic and conditional
utility maximization problem in incomplete markets for utility
functions defined on the whole
real line and establish relations between optimal strategies of these problems.

To this end we first give some definitions and auxiliary assertions.

We shall say that an adapted stochastic process
$(X_t, t\in[\tau,T])$
is a generalized martingale (resp. supermartingale) if

1) $E(|X_t|/F_\tau)<\infty$, for any $t\in[\tau,T]$

2) $E(X_t/F_{t'})=X_{t'}$\;\;\;(resp. $\le X_{t'})$ for any
$t'\le t$, where $t',t\in[\tau,T]$

( see the definition of generalized
conditional expectations and of generalized supermartingales for
discrete time in \cite{Shr})

{\bf Definition.} A predictable $S$ integrable process
$\pi$ is in $\Pi_{x,\tau}$, if $E(U(x+\int_\tau^T\pi_udS_u)/F_\tau)$
is finite and $((\pi\cdot S)_t, t\ge\tau)$ is
a generalized supermartingale
under each $Q\in{\cal M}^a$ with finite $\wt U$-expectation
$E\wt U(\frac{dQ}{dP})$.

We shall also need two complementary assumptions
 \begin{assumption}\label{as2}
The filtration $F$ is continuous
and $\liminf\limits_{y\to\infty} Z_T(y)/y>0$ for the process $Z_T(y)=y\frac{dQ^*(y)}{dP}=y\rho_T^*(y)$.
\end{assumption}
\begin{assumption}\label{as2}
The utility function $U$ is two times differentiable
and there are
constants $c_1>0$ and $c_2>0$ such that
\begin{equation}\label{rav1}
   c_1< -\frac{U^{''}(x)}{U^{'}(x)}<c_2, \;\; x\in R.
\end{equation}
\end{assumption}
The last condition is similar to the condition
on relative risk-aversion introduced in \cite{KS}. Note that for exponential utility function
the risk-aversion coefficient $-\frac{U^{''}(x)}{U^{'}(x)}=\gamma$ is a constant and
condition (\ref{rav1}) is also satisfied for linear combinations of exponential utility functions with different risk-aversion parameters.

The proof of the following assertion follows from Theorem 4.1 and
Proposition 3.1 of \cite{MT15}.

{\bf Proposition 1.} Let Assumptions 1-3 be satisfied.

Then for any $t\in [0,T]$ there exists a modification of the
optimal wealth process $(X_t(x), x\in R)$ (resp. of $Z_t(y)$)
almost all paths of
which are strictly increasing  and absolutely continuous with
respect to $dx$ (resp. $dy$). Besides
\begin{equation}\label{ax4}
 X'_t(x)>0,\;\;\;\;  E^{Q(x)}(X'_T(x))^2\le C,
 \end{equation}
 \begin{equation}\label{infty}
  \lim_{x\to\infty}X_t(x)=\infty,\;\;\;
    \lim_{x\to -\infty}X_t(x)=-\infty
\end{equation}
$P$-a.s. for any $t\in [0,T]$ and the adapted inverse
$X_t^{-1}(x)$ (resp. $Z_t^{-1}(y)$) of the optimal wealth process
exists.

We shall need also the continuity properties of  the square characteristics
$\la X(x)-X(y)\ra$ which can be deduced from Proposition 1.

{\bf Lemma 1.} Let conditions of Proposition 1 be satisfied.
Then, for any $t\in [0,T]$ the random field
 $(\la X(x)-X(y)\ra_t, x,y\in R)$  admits a continuous modification.

{\it Proof.}
It follows from Proposition 1 that $X_t(b)-X_t(a)=
\int_a^bX_t'(x)dx$ and
$$
\int_a^bE^{Q(x)}\la X'(x)\ra_Tdx=
\int_a^bE^{Q(x)}X_T'(x)^2dx<\ift
$$
and by the Fubini theorem $\int_a^b\frac{U'(X_T(x))}{V'(x)}\la X'(x)\ra_Tdx<\ift,\;P-a.s.$
Thus by continuity of
$\frac{V'(x)}{U'(X_T(x))}$
we obtain
$$
\int_a^b\la X'(x)\ra_Tdx\le\max_{x\in[a,b]}\frac{V'(x)}{U'(X_T(x))}\int_a^b\frac{U'(X_T(x))}
{V'(x)}\la X'(x)\ra_Tdx<\ift,\;P-a.s.
$$
Therefore, using the Kunita-Watanabe and H\"older's inequalities we have
\begin{align*}
\la X(b)-X(a)\ra_t
=\int_a^b\int_a^b\la X'(x),X'(y)\ra_tdxdy\le\\
\le\int_a^b\int_a^b\la X'(x)\ra_t^{1/2}\la X'(y)\ra_t^{1/2}dxdy
=\left(\int_a^b\la X'(x)\ra_t^{1/2}dx\right)^2\le\\
\le (b-a)\int_a^b\la X'(x)\ra_t dx<\ift,\;P-a.s.
\end{align*}
and it follows from inequality
\begin{align*}
\la X(b')-X(a')\ra_t-\la X(b)-X(a)\ra_t
\\
\le\la X(b')-X(b)\ra_t^{1/2}\la X(b')-X(a')+X(b)-X(a)\ra_t^{1/2}\\
+\la X(a')-X(a)\ra_t^{1/2}\la X(b')-X(a')+X(b)-X(a)\ra_t^{1/2}
\end{align*}
that  $\la X(b_n)-X(a_n)\ra_t\to \la X(b)-X(a)\ra_t,\;P-a.s.$ when $b_n\to b,\;a_n\to a$.
Thus the stochastic field defined by
$$
\la X(x)-X(y)\ra_t^*=\left\{
                       \begin{array}{ll}
                         \lim_{r\to a,r'\to b}\la X(r)-X(r')\ra_t, & {r,r'\;are\;rational}, \\
                         0, & {if\; the\;limit\; does\; not\; exists}
                       \end{array}
                     \right.
$$
is continuous and stochastically equivalent to $\la X(x)-X(y)\ra_t$.

 {\bf Theorem 1.} Let Assumptions 1-3 be satisfied.
 Then there exist the maximizer of (\ref{ut5}) and the minimizer of
 (\ref{utd}) in the classes
$\Pi_{\tau,x}$ and ${\cal Z}_{\tau,y}$ respectively and equalities
\beq\label{prim}
X_T(\tau,x)=X_T(X_\tau^{-1}(x)),\;\pi_t(\tau,x)=\pi_t(X_\tau^{-1}(x)),t\ge\tau,
\eeq
\beq\label{dual}
Y(\tau,y)=Z_T(Z_\tau^{-1}(y)),\;\rho_T^{Q^*}(\tau,y)=\rho_\tau^{Q^*}(y)\frac{Z_T(Z_\tau^{-1}(y))}{y}
\eeq
are satisfied.

Moreover $P$-a.s.
\begin{equation}\label{ax8}
V(\tau,x)=E\bigg[U\bigg(x+\int_\tau^T\pi_u(X_\tau^{-1}(x))dS_u\bigg)\mid {F}_\tau\bigg],\\
\end{equation}
$$
\wt V(\tau,y)
=E\Big[\widetilde U(Z_T(Z_\tau^{-1}(y)))\mid F_\tau\Big],
$$
the following duality relation holds
\begin{equation}\label{ax9}
U'\left(x+\int_\tau^T\pi_u(X_\tau^{-1}(x))dS_u\right)=Z_T(Z_\tau^{-1}(y)),\;y=V'(\tau,x)
\end{equation}
and the process
\begin{equation}\label{marti}
Z_t(Z^{-1}_\tau(y))X_t(X_\tau^{-1}(x)),\;\;\;t\in[\tau,T],\;\;where\;\;\;y=V'(\tau,x),
\end{equation}
is a generalized martingale.

{\it Proof.}
By the optimality principle (see, e.g. \cite{MT10}) $V(t,X_t(x))$
is a martingale and since $V(T,x)=U(x)$ we have that for any $x\in R$
\begin{equation}\label{ax6}
V(\tau,X_\tau(x))=
E\big(U(X_T(x))/{\mathcal F}_\tau\big) \;\;\; P-a.s.
\end{equation}
Since for any $\tau$ the functions $V(\tau, x)$ and $X_\tau(x)$ are continuous
for almost all $\omega\in\Omega$, the equality
(\ref{ax6})
holds $P$-a.s. for all $x\in R$ and
substituting $X_\tau^{-1}(x)$ in this equality
we obtain that
$$
V(\tau,x)= E\big(U(X_T(X_\tau^{-1}(x)))/F_\tau\big) \;\;\; P-a.s.,
$$
 which means the maximality of $X_T(X_\tau^{-1}(x))$.
 Let us show that $X_T(X_\tau^{-1}(x))$ is equal to the stochastic
 integral
 \begin{equation}\label{ax7}
 X_T(X_\tau^{-1}(x))=x+\int_\tau^T\pi_u(X_\tau^{-1}(x))dS_u
 \end{equation}
 and that   $\pi(X_\tau^{-1}(x))$ belongs to the class $\Pi_{\tau,x}$

In order to show equality (\ref{ax7}) it is enough to show that
$\int_\tau^T\pi_u(x)dS_u\big|_{x=\xi}=\int_\tau^T\pi_u(\xi)dS_u,$
for $\xi=X_\tau^{-1}(x)$.

Let us consider the sequence of simple random variables
$\xi_n=\sum_{k=-\ift}^\ift c_k1_{A_k},$ where $A_k=(\frac{k}{n}\le\xi<\frac{k+1}{n}),\;c_k=\frac{k}{n}$. We have
$\xi_n\to\xi$ uniformly and
$$
\int_\tau^T\pi_u(\xi_n)dS_u=\sum_{k=-\ift}^\ift \int_\tau^T\pi_u(c_k)1_{A_k}dS_u=
$$
$$
=\sum_{k=-\ift}^\ift 1_{A_k}\int_\tau^T\pi_u(c_k)dS_u=\int_\tau^T\pi_u(x)dS_u\big|_{x=\xi_n}.
$$
On the other hand
$$
\int_\tau^T\pi_u(x)dS_u\big|_{x=\xi_n}-\int_\tau^T\pi_u(x)dS_u\big|_{x=
\xi}=
$$
$$
=X_T(\xi_n)-X_\tau(\xi_n)-(X_T(\xi)-X_\tau(\xi))\to 0,
$$
as $n\to\ift$, since $X_t(x)$ is continuous and
$$
\int_\tau^T(\pi_u(\xi_n)-\pi_u(\xi))^2d\la S\ra_u=
$$
$$
=\la X(x)-X(y)\ra_T-\la X(x)-X(y)\ra_\tau|_{x=\xi_n,y=\xi}\to 0,P-a.s.
$$
as $n\to\ift$, by continuity of $\la X(x)-X(y)\ra_t$. Hence $\int_\tau^T\pi_u(\xi_n)dS_u\to \int_\tau^T\pi_u(\xi)dS_u$ in probability and
$\int_\tau^T\pi_u(x)dS_u\big|_{x=\xi}=\int_\tau^T\pi_u(\xi)dS_u-P.a.s..$

Since $E|U(X_{T}(x))|<\infty $ and $E^Q|X_t(x)|<\infty, t\in[0,T]$ for any
$Q\in{\cal M}^a$ and $X^{-1}_\tau(x)$ is $F_\tau$-measurable we have that
$$
E[|U(X_{T}(X_\tau^{-1}(x)))|\mid{F}_\tau]<\infty,\;\;\;
E^Q(|X_t(X^{-1}_\tau(x))|/F_\tau)<\infty\;\;\; P- a.s.,  t\ge\tau
$$
On the other hand, since for any $t\in [0,T]$ the function $(X_t(x), x\in R)$  is
continuous and increasing, the supermartingale inequality
$E^Q(X_t(x)/F_{t'})\le X_{t'}(x), \;\;\;t'\le t\le T$ implies
that
$$
E^Q(X_t(X^{-1}_\tau(x))/F_{t'})\le X_{t'}(X^{-1}_\tau(x)),\;\;\;
\tau\le t'\le t\le T
$$
for any  $Q\in{\cal M}^a$, hence $\pi(\tau, x)=\pi(X^{-1}_\tau(x))$
belongs
to the class $\Pi_{\tau,x}$ and the equality (\ref{ax8}) holds.
Similarly one can show the minimality of $Z_T(Z_\tau^{-1}(y))$, so
conditional density of the minimax martingale measure
to the problem (\ref{ut5}) is $\frac{Z_T(Z_\tau^{-1}(y))}{y}$.

Since for any $t\in[0,T]$ the functions $V'(t,x), x\in R$ and
$Z_t(y), y>0$ are continuous and the inverse of $Z_t(y)$ exists,
from (\ref{main2}) we
have that  $P$-a.s.
\begin{equation}\label{ax10}
Z^{-1}_\tau(V'(\tau,x))=V'(X^{-1}_\tau(x))
\end{equation}
which, together with (\ref{main}) implies
the conditional duality relation (\ref{ax9}).

Note also that since $Z_t(y)X_t(x)$ is a martingale (see Theorem 1 from \cite{S3}), by continuity of
$X(x)$ and $Z(y)$  the process
$(Z_t(V'(X^{-1}_\tau))X_t(X^{-1}_\tau(x)), t\ge\tau)$  will be
a generalized martingale and by equality (\ref{ax10}) this is equivalent to (\ref{marti}).
\qed

\section{Relations between decomposition terms of the value processes of primal and dual problems}

In this section additionally to the continuity of the filtration $F$ we assume that any orthogonal to $M$ local martingale $L$ is represented as a stochastic integral with respect to the given
continuous local martingale $M^\bot$. Therefore, the value process $V(t,x)$ admits the decomposition
$$
V(t,x)=V(0,x)-A(t,x)+\int_0^t\varphi(s,x)dM_s+\int_0^t\varphi_\bot(s,x)dM^\bot_s,
$$
where $ A(t,x)$ is an increasing process for any $x\in R$ , ${\varphi}$ and ${\varphi}^\bot$
are $M$ and ${M^\bot}$ integrable predictable processes respectively.
Since the value process
${\widetilde V}(t,y)$  of the dual problem is a submartingale for each $y>0$ it is decomposable as
\begin{equation}\label{eqdec}
{\widetilde V}(t,y)=\widetilde V(0,y)+\widetilde A(t,y)+\int_0^t\widetilde\varphi(s,y)dM_s+\int_0^t{\widetilde\varphi}_\bot(s,y)dM^\bot_s,
\end{equation}
with $M$ and ${M^\bot}$ integrable predictable processes $\widetilde{\varphi}$ and ${\widetilde\varphi}^\bot$ and an increasing process ${\widetilde A}(t,y)$.

It is known that the value processes of the primal and dual problems are related by the equality
\begin{equation}\label{eqrel}
V(t,-{\widetilde V}'(t,y))={\widetilde V}(t,y)-y{\widetilde V}'(t,y).
\end{equation}
We are interested how are related the decomposition terms $A,\varphi$ and $ \varphi_\bot$ with
$\widetilde A, \widetilde\varphi$ and $\widetilde\varphi_\bot$ respectively.

{\bf Theorem 2.} Assume that the filtration $F$ is continuous and   any orthogonal to $M$ local martingale $L$ is represented as a stochastic integral with respect to  a local martingale $M^\bot$.
Assume that $V(t,x)$ is a regular family of semimartingales (i.e., satisfies conditions a)-c) of introduction) and that ${\widetilde V}'(t,y)$ is a semimartiongale with the decomposition
\begin{equation}\label{eqder}
{\widetilde V}'(t,y)=\widetilde V'(0,y)+\widetilde B(t,y)+\int_0^t\widetilde\varphi'(s,y)dM_s+\int_0^t{\widetilde\varphi}'_\bot(s,y)dM^\bot_s,
\end{equation}
where $\widetilde B(t,y)$ is the process of finite variation for any $y$.

 Then $({\widetilde V}(t,y), y>0)$ is a regular family of semimartingales and
\begin{equation}\label{eqm1}
\widetilde\varphi(s, y)=\varphi(s,-{\widetilde V}'(s,y)),\;\;\;\mu^{\la M\ra}\;\;a.e.,
\end{equation}
\begin{equation}\label{eqm2}
\widetilde\varphi_\bot(s, y)=\varphi_\bot(s,-{\widetilde V}'(s,y)),\;\;\;\mu^{\la M\ra}\;\;a.e.,
\end{equation}
$$
\widetilde A(t,y)=\int_0^ta(s,-\widetilde{V}'(s,y))d\la M\ra_s-\frac{1}{2}\int_0^t\frac{(\varphi'(s,-{\widetilde V}'(s,y)))^2}{V''(s, -{\widetilde V}'(s,y))}d\la M\ra_s-
$$

\begin{equation}\label{eqa1}
- \frac{1}{2}\int_0^t\frac{(\varphi'_\bot(s,-{\widetilde V}'(s,y)))^2}{V''(s, -{\widetilde V}'(s,y))}d\la M^\bot\ra_s.
\end{equation}
Besides $\widetilde V(t,y)$ satisfies the  BSPDE
$$
\widetilde V(t,y)=\widetilde V(0,y)+\int_0^t\big(y\lambda_s\widetilde\varphi'(s,y)-\frac{1}{2}y^2\lambda^2_s\widetilde V''(s,y)\big)
d\la M\ra_s
$$
\begin{equation}\label{eqpded}
 +\frac{1}{2}\int_0^t\frac{(\varphi'_\bot(s,y))^2}{\widetilde V''(s,y)}d\la M^\bot\ra_s+\int_0^t\widetilde\varphi(s,y)dM_s+\int_0^t\widetilde\varphi_\bot(s,y)dM^\bot_s,\;\;
\widetilde V(T,y)=\widetilde U(y).
\end{equation}

{\it Proof.} Using  the duality relation (\ref{eqrel})
and  the It\^o-Ventzel formula (see, e.g., \cite{K} or \cite{V})  we have
$$
V(t,-\widetilde{V}'(t,y))=
$$
$$
=V(0,-\widetilde{V}'(0,y))+\int_0^t\varphi(s,-\widetilde{V}'(s,y))dM_s+
\int_0^t\varphi_\bot(s,-\widetilde{V}'(s,y))dM^\bot_s-
$$
$$
-\int_0^tV'(s,-\widetilde{V}'(s,y))\widetilde\varphi'(s,y)dM_s
-\int_0^tV'(s,-\widetilde{V}'(s,y))\widetilde\varphi'_\bot(s,y)dM^\bot_s-
$$
$$
+ \int_0^ta(s,-\widetilde{V}'(s,y))d\la M\ra_s-\int_0^tV'(s,-\widetilde{V}'(s,y))d\widetilde B(s,y)+
$$
$$
+\frac{1}{2}\int_0^tV''(s,-\widetilde{V}'(s,y))\widetilde\varphi'(s,y)^2d\la M\ra_s+
$$
$$
+\frac{1}{2}\int_0^tV''(s,-\widetilde{V}'(s,y))\widetilde\varphi_\bot'(s,y)^2d\la M^\bot\ra_s
$$
$$
-\int_0^t\varphi'(s,-\widetilde{V}'(s,y))\widetilde\varphi'(s,y)d\la M\ra_s-
$$
$$
-\int_0^t\varphi_\bot'(s,-\widetilde{V}'(s,y))\widetilde\varphi_\bot'(s,y)d\la M^\bot\ra_s=
$$
$$
=\widetilde A(t,y)+\int_0^t\widetilde\varphi(s,y)dM_s+\int_0^t{\widetilde\varphi}_\bot(s,y)dM^\bot_s-
$$
\begin{equation}\label{eq20}
-y\widetilde B(t,y)-y\int_0^t\widetilde\varphi'(s,y)dM_s-y\int_0^t{\widetilde\varphi}'_\bot(s,y)dM^\bot_s.
\end{equation}

Since $V'(s,-\widetilde{V}'(s,y))=y$, from (\ref{eq20}) we obtain that
$$
\int_0^t\varphi(s,-\widetilde{V}'(s,y))dM_s+
\int_0^t\varphi_\bot(s,-\widetilde{V}'(s,y))dM^\bot_s
$$
$$
+ \int_0^ta(s,-\widetilde{V}'(s,y))d\la M\ra_s+\frac{1}{2}\int_0^tV''(s,-\widetilde{V}'(s,y))(\widetilde\varphi'(s,y))^2d\la M\ra_s+
$$
$$
+\frac{1}{2}\int_0^tV''(s,-\widetilde{V}'(s,y))(\widetilde\varphi_\bot'(s,y))^2d\la M^\bot\ra_s
$$
$$
-\int_0^t\varphi'(s,-\widetilde{V}'(s,y))\widetilde\varphi'(s, y))d\la M\ra_s-
$$
$$
-\int_0^t\varphi_\bot'(s,-\widetilde{V}'(s,y))\widetilde\varphi_\bot'(s, y))d\la M^\bot\ra_s=
$$
\begin{equation}\label{eq21}
=\widetilde A(t,y)+\int_0^t\widetilde\varphi(s,y)dM_s+\int_0^t\widetilde\varphi_\bot(s,y)dM^\bot_s.
\end{equation}
Equalizing the martingale parts in (\ref{eq21}) we obtain equalities  (\ref{eqm1})  and (\ref{eqm2}).
Since $\widetilde V(t,y)$ is two-times differentiable and
\begin{equation}\label{eq23}
\widetilde V''(t,y)=-\frac{1}{V''(t,-\widetilde V'(t,y))},
\end{equation}
we have that  $\widetilde\varphi(s,y)$  and $\widetilde\varphi_\bot(s,y)$ are also differentiable and
\begin{equation}\label{eqmd1}
\widetilde\varphi'(s, y)=\varphi'(s,-\widetilde V'(s,y))\widetilde V''(t,y)=\frac{ \varphi'(s,-\widetilde V'(s,y))}{V''(t,-\widetilde V'(s,y))} ,\;\;\;\mu^{\la M\ra}\;\;a.e.,
\end{equation}
\begin{equation}\label{eqmd2}
\widetilde\varphi_\bot'(s, y)=\varphi'_\bot(s,-\widetilde V'(s,y))\widetilde V''(t,y)=\frac{ \varphi_\bot'(s,-\widetilde V'(s,y))}{V''(t,-\widetilde V'(s,y))},\;\;\;\mu^{\la M^\bot\ra}\;\;a.e.
\end{equation}
Therefore,
$$
\varphi'(s,-\widetilde{V}'(s,y))\widetilde\varphi'(s,y))=V''(s,-\widetilde{V}'(s,y))(\widetilde\varphi'(s,y))^2 ,\;\;\;\mu^{\la M\ra}\;\;a.e.,
$$
$$
\varphi'_\bot(s,-\widetilde{V}'(s,y))\widetilde\varphi'_\bot(s,y)=V''(s,-\widetilde{V}'(s,y))(\widetilde\varphi'_\bot(s,y))^2 ,\;\;\;\mu^{\la M^\bot\ra}\;\;a.e.
$$
and equalizing the finite variation parts in  (\ref{eq21}) we deduce that equality  (\ref{eqa1}) holds.

Let us show now that $\widetilde V(t,y)$ satisfies the BSPDE (\ref{eqpded}).
It follows from (\ref{ut6}) that

$$
a(s,x)=\frac{1}{2} \frac{(\lambda_sV'(s,x)+\varphi'(s,x))^2}{V''(s,x)}.
$$
Therefore, using equalities  $V'(s,-\widetilde{V}'(s,y))=y$ ,  (\ref{eq23}) and (\ref{eqmd1})
$$
\int_0^ta(s,-\widetilde{V}'(s,y))d\la M\ra_s=
\frac{1}{2} \int_0^t\frac{(y\lambda_s+\varphi'(s,-\widetilde{V}'(s,y)))^2}{V''(s-,\widetilde{V}'(s,y))}d\la M\ra_s=
$$
$$
=\int_0^t\big(y\lambda_s\widetilde\varphi'(s,y)-\frac{1}{2}y^2\lambda^2_s\widetilde V''(s,y)\big)
d\la M\ra_s
+\frac{1}{2}\int_0^t\frac{(\varphi'(s,-{\widetilde V}'(s,y)))^2}{V''(s, -{\widetilde V}'(s,y))}d\la M\ra_s.
$$
which (together with (\ref{eqa1})) implies that
$$
\widetilde A(t,y)=\int_0^t\big(y\lambda_s\widetilde\varphi'(s,y)-\frac{1}{2}y^2\lambda^2_s\widetilde V''(s,y)\big)
d\la M\ra_s
$$
\begin{equation}\label{eqd}
+\frac{1}{2}\int_0^t\frac{(\varphi'_\bot(s,y))^2}{\widetilde V''(s,y)}d\la M^\bot\ra_s.
\end{equation}
Now, (\ref{eqdec}) and (\ref{eqd}) imply that $\widetilde V(t,y)$ satisfies (\ref{eqpded}).

{\bf Remark.} It follows from (\ref{eqdec}), (\ref{eqa1}) and (\ref{eq23}) that
$\widetilde V(t,y)$ satisfies also the forward SPDE derived in  \cite{El}, which takes in this case the
following form
$$
\widetilde V(t,y)=\int_0^t a(s,-\widetilde{V}'(s,y))d\la M\ra_s+
\frac{1}{2}\int_0^t(\varphi'(s,-\widetilde{V}'(s,y))^2\widetilde{V}''(s,y)d\la M\ra_s+
$$
$$
\frac{1}{2}\int_0^t(\varphi'_\bot(s,-\widetilde V'(s,y)))^2\widetilde V''(s,y)d\la M^\bot\ra_s+
$$
$$
 +\int_0^t\widetilde\varphi(s,-\widetilde{V}'(s,y))dM_s+\int_0^t\widetilde\varphi_\bot(s,-\widetilde{V}'(s,y))dM^\bot_s
$$

\section{Differential equation for the inverse flow of the optimal wealth}

By Proposition 1, if the filtration $F$ is continuous and Assumptions 1-3 are satisfied
then  the adapted inverse $X_t^{-1}(x)$ of the optimal wealth process exists.
Under stronger conditions we shall derive  for the inverse  process $X_t^{-1}(x)$ a Stochastic Differential Equation (SDE) which will be used
 to show the absolutely continuity of
bounded variation parts of $V(t,x)$ and $V'(t,x)$ with respect to  square characteristic $<S>$.

For stochastic process $\xi_t(x)$  by $\xi'_t(x)$ (or $\partial \xi_t(x)$) we denote the derivative with respect to $x$,
$\mu^{\la S\ra}$ denotes  Dolean's measure for $\la S\ra$, i.e. the measure $d\la S\ra dP$ on $[0,T]\times \Omega$.
If $F(t,x)$ is a family of semimartigales   then $\int_0^TF(ds,\xi_s)$ denotes a generalized stochastic integral (see \cite{K}),
or stochastic line integral by terminology from \cite{Ch}.
If $F(t,x)=xG_t$, where $G_t$ is a semimartingale then  the generalized stochastic integral coincides with usual one
denoted by $\int_0^T\xi_sdG_s$ or $(\xi\cdot G)_T$.

Now we shall derive an SDE for the inverse of the optimal wealth $\psi_t(x)=X_t^{-1}(x)$ of the form
\beq\lbl{sd1}
d\psi_t=\sg_t(\psi_t)dS_t+\mu_t(\psi_t)d\la S\ra_t,\;\psi_0=x,
\eeq
where $\sg_t(z)=-\frac{\pi_t(z)}{X'_t(z)},\;\mu_t(z)=\frac{1}{2X'_t(z)}\left(\frac{\pi_t^2(z)}{X'_t(z)}\right)'$.

{\bf Proposition 2.}
Let  $X''_t(x),\;\pi'_t(x)$  exist  $\mu^{\la S\ra}$-a.e. and are locally Lipschitz functions with respect to $x$  $\mu^{\la S\ra}-$a.e..
Then SDE (\ref{sd1}) or equivalently
\beq\lbl{sd}
d\psi_t=-\frac{\pi_t(\psi_t)}{X'_t(\psi_t)}dS_t+\frac{\pi'_t(\psi_t)\pi_t(\psi_t)}{X'_t(\psi_t)^2}d\la S\ra_t-\frac{1}{2}\frac{X''_t(\psi_t)\pi_t^2(\psi_t)}{X'_t(\psi_t)^3}d\la S\ra_t,
\eeq
\beq
\psi_0=x
\eeq
admits a unique maximal solution and it coincides with $X_t^{-1}(x)$.

{\it Proof}. The SDE (\ref{sd1}) admits unique maximal solution up to time $\tau(x)\le T$, where $|\psi_{\tau(x)-}|=\infty$ if $\tau(x)<T$
(see \cite{K}).
Applying the Ito-Ventzel formula for $X_t(\psi_t)\equiv X(t,\psi_t)$ (see \cite{K} or \cite{V})
and using that $\psi_t$ satisfies (\ref{sd}) we get
\begin{eqnarray*}
  dX(t,\psi_t) = X(dt,\psi_t)+X'(t,\psi_t)d\psi_t&+&\frac{1}{2}X''(t,\psi_t)d\la\psi\ra_t \\
  +d\left\la \int_0^\cdot X'(dr,\psi_r(x)),\psi(x)\right\ra_t= \pi_t(\psi_t)dS_t&+&
  X'_t(\psi_t)[-\frac{\pi_t(\psi_t)}{X'_t(\psi_t)}dS_t\\
  +\frac{\pi'_t(\psi_t)\pi_t(\psi_t)}{X'_t(\psi_t)^2}d\la S\ra_t&-&\frac{1}{2}\frac{X''_t(x)\pi_t^2(\psi_t)}{X'_t(\psi_t)^3}d\la S\ra_t]
  \\
  +\frac{1}{2}\frac{X''_t(x)\pi_t^2(\psi_t)}{X'_t(\psi_t)^2}d\la S\ra_t&-&\frac{\pi'_t(\psi_t)\pi_t(\psi_t)}{X'_t(\psi_t)}d\la S\ra_t=0,\\
  \psi_0(x)&=&x.
\end{eqnarray*}
Hence $X(t,\psi_t(x))=x$ on $[0,\tau(x))$ .   Since    $|X_{\tau(x)}^{-1}(x)|<\infty,$ we have $\tau(x)=T$ $ P-$a.s.
and $\psi_t(x)=X_t^{-1}(x).$       \qed
\be{rem}
Let $\pi_t(x)=H_t(X_t(x))$. Then
$$
d\psi_t=-\frac{H_t(X_t(\psi_t))}{X'_t(\psi_t)}dS_t+\frac{H'_t(X_t(\psi_t))H_t(X_t(\psi_t))}{X'_t(\psi_t)^2}d\la S\ra_t-\frac{1}{2}\frac{X''_t(\psi_t)H_t^2(X_t(\psi_t))}{X'_t(\psi_t)^3}d\la S\ra_t.
$$
Using equalities $X_t(\psi_t(x))=x,\;\frac{1}{X'_t(\psi_t(x))}=\psi_t'(x),\;and\;-\frac{X''_t(\psi_t(x))}{X'_t(\psi_t(x))}=\frac{\psi_t''(x)}{\psi_t'(x)^2}$
we obtain the linear Partial SDE (linear PSDE)
$$
d\psi_t(x)=-H_t(x)\psi'_t(x)dS_t+H'_t(x)H_t(x)\psi'_t(x)d\la S\ra_t+\frac{1}{2}H_t^2(x)\psi''_t(x)d\la S\ra_t
$$
or a PSDE in the divergence form
$$
d\psi_t(x)=-H_t(x)\psi'_t(x)dS_t+\frac{1}{2}(H_t^2(x)\psi'_t(x))'d\la S\ra_t.
$$
\ee{rem}
Let us define martingale random fields
\begin{eqnarray}
      \nonumber {\cal M}(t,x)=E[U(X_T(x)|F_t], \\
     \nonumber \overline{\cal M}(t,x)=E[U'(X_T(x)|F_t].
   \end{eqnarray}

{\bf Proposition 3.}
Let   conditions of Proposition 2 be satisfied.
\begin{description}
     \item{i)}
     If $\mathcal{M}(t,x)$ is two times continuously differentiable with respect to $x$, then the finite variation part of $V(t,x)=\mathcal{M}(t,\psi_t(x))$
is absolutely continuous with respect to $\la S\ra$.
     \item{ii)} If $\overline{\cal M}(t,x)$ is two times continuously differentiable with respect to $x$, then $V^{'}(t,x)$ is a special semimartingale and
the finite variation part of $V'(t,x)={\overline{\mathcal M}}(t,\psi_t(x))$
is absolutely continuous with respect to $\la S\ra$. Besides $V'(t,x)$ admits the decomposition
   \end{description}
\begin{equation}\label{fin}
V'(t,x)=V'(0, x)- \int_0^ta'(s,x)\,d\langle M\rangle_s+\int_0^t\psi'(s,x)\,dM_s+L'(t,x).
 \end{equation}
{\it Proof}.
i)
By the optimality principle $V(t,X_t(x))$ is a martingale and since $V(T,x)=U(x)$ we have that
$V(t,X_t(x))=E[U(X_T(x))|{F}_t]={\cal M}(t,x)$.
Therefore by duality relation (\ref{main})
\begin{equation}\label{bar}
{\cal M}'(t,x)=V'(t,X_t(x))X_t'(x)=Z_t(y)X_t'(x)
\end{equation}
is a martingale and let
$$
{\cal M}'(t,x)=V'(x)+\int_0^th_r(x)dM_r+L_t(x),\;\;\;L(x)\bot M
$$
be  the GKW  decomposition
of ${\cal M}'(t,x)$.
From (\ref{sd})
we have
\beq\lbl{line}
\left\la \int_0^\cdot\mathcal{M'}(dr,\psi_r(x)),\psi(x)\right\ra_t=-\int_0^th_r(\psi_r(x))\frac{\pi_r(\psi_r(x))}
{X'_r(\psi_r(x))}d\la S\ra_r .
\eeq
Since $V(t,x)={\cal M}(t,X_t^{-1}(x))$, by the Ito-Ventzel formula we get
\begin{eqnarray}\lbl{12}
     V(t,x)=V(0,x)+\int_0^t{\cal M}(ds,\psi_s)&+&\int_0^t{\cal M}'(s,\psi_s)d\psi_s \\
     \nonumber +\frac{1}{2}\int_0^t{\cal M}''(s,\psi_s)d\la\psi\ra_s&+&\left\la\int_0^\cdot\mathcal{M'}(dr,\psi_r(x)),\psi(x)\right\ra_t
   \end{eqnarray}
In view of (\ref{sd}) and (\ref{line}) one can verify that all finite variation members of (\ref{12}) are integrals with respect to $\la S\ra$.
Namely,
\begin{multline*}
  -A(t,x)\\
  =\int_0^t{\cal M}'(r,\psi_r(x))\left(\frac{\pi'_r(\psi_r(x))\pi_r(\psi_r(x))}{X'_r(\psi_r(x))^2}-\frac{1}{2}\frac{X''_r(\psi_r(x))\pi_r^2(\psi_r(x))}{X'_r(\psi_r(x))^3}\right)d\la S\ra_r
  \\+\int_0^t\left(\frac{1}{2}{\cal M}''(r,\psi_r(x))\frac{\pi_r^2(\psi_r(x))}{X'_r(\psi_r(x))^2}-h_r(\psi_r(x))\frac{\pi_r(\psi_r(x))}
{X'_r(\psi_r(x))}\right)d\la S\ra_r.
\end{multline*}
ii) It follows from (\ref{main}) and (\ref{main2}) that
\beq\lbl{mart}
\overline{\cal M}(t,x)=E[U'(X_T(x))|F_t]=E[Z_T(y)|F_t]=Z_t(y)=V'(t,X_t(x)),
\eeq
which (together with (\ref{bar})) implies that $\cal M$ and $\overline{\cal M}$ are related as
   \begin{equation}\label{calm}
   {\mathcal M'}(t,x)={\overline{\cal M}}(t,x)X'_t(x)
   \end{equation}
and $V'(t,x)=\overline{\cal M}(t,X_t^{-1}(x)).$ It follows from (\ref{mart}) that $\overline{\cal M}'(t,x)=Z_t'(y)V''(x)$ is a
martingale and
\beq\lbl{line1}
\left\la \int_0^\cdot\overline{\cal M}'(dr,\psi_r(x)),\psi(x)\right\ra_t=-\int_0^t{\bar h}_r(\psi_r(x))\frac{\pi_r(\psi_r(x))}
{X'_r(\psi_r(x))}d\la S\ra_r,
\eeq
where $\overline{\cal M}'(t,x)=\bar V''(x)+\int_0^t{\bar h}_r(x)dM_r+{\bar L}_t(x),\;{\bar L}(x)\bot M$ is
the GKW decomposition
of $\overline{\cal M}'(t,x)$.
Therefore the Ito-Ventzel formula implies that $V^{'}(t,x)={\overline{\cal M}}(t, X_t^{-1}(x))$ is a special semimartingale and similarly to i)
one can show that the finite variation part of $V^{'}(t,x)$ is absolutely continuous with respect to $\la S\ra$. Therefore,
$V'(t,x)$ is decomposable as
\beq\lbl{value'}
V'(t,x)=V'(0,x)+\int_0^tb(r,x)d\la M\ra_r+\int_0^tg(r,x)dM_r+N(t,x),
\eeq
for some local martingale $N(t,x)$ orthogonal to $M$ for any $x\in R$ and  $M$ and $\la M\ra$ integrable processes $g$ and $b$ respectively.
The It\^o-Ventzel formula and conditions of this proposition also imply that $b(r,x)$ and  $ g(r, x)$  are continuous at $x$. Therefore,
integrating the  equation (\ref{value'}) with respect to $dx$ (over a finite interval)  and using the stochastic Fubini theorem (taking decomposition (\ref{value}) in mind),
 we obtain  (\ref{fin}).
\qed

\section{The case of complete markets}

In this section for the case of complete markets we provide sufficient conditions on the utility function $U$  which guarantee an existence of
a solution of BSPDE (\ref{ut6}).

Hereafter we shall assume that the market is complete, i.e.
  $$
  dQ=Z_TdP,\;\;\;  {\text where }\;\;\; Z_T=\mathcal{E}_T(-\ld\cdot M)
  $$
  is the unique martingale measure.
Let
\begin{equation}\label{rav}
R_1(x)=-\frac{U^{''}(x)}{U^{'}(x)}, \quad R_2(x)=-\frac{U^{'''}(x)}{U^{''}(x)},\;\;\;\;x\in R.
\end{equation}

We shall use one of the following conditions:
\begin{description}
  \item{r1}) $U$ is  three-times differentiable, $R_1(x)$ is bounded away from zero and infinity and $R_2(x)$ is bounded and Lipschitz continuous.

  \item{r2}) $U$ is four-times differentiable and the density  $Z_T$ of the unique martingale measure is bounded.
\end{description}

{\bf Lemma 2}.  Let the market be complete and  condition r1) be satisfied.
Then the optimal wealth $X_T(x)$ is two-times
differentiable and  the derivatives $X_T^{'}(x), X_T^{''}(x)$ are bounded and Lipschitz continuous.

{\it Proof.} Since ${\wt U}(y)$ and $U(x)$ are conjugate, ${\wt U}(y)$ is also three-times differentiable and
\begin{equation}\label{sami}
\wt U^{''}(y)=-\frac{1}{U^{''}(x)},\;\;\;\; \wt U^{'''}(y)=-\frac{U^{'''}(x)}{(U^{''}(x))^3}, \;\;\;y=U^{'}(x).
\end{equation}
Therefore the functions $B_1(y)$ and $B_2(y)$, where
\begin{equation}\label{b1}
B_1(y)=y{\wt U}^{''}(y)=1/R_1(x), \;\;\;\;
B_2(y)=y^2{\wt U}^{'''}(y)=R_2(x)/R_1^2(x)
\end{equation}
respectively, are also bounded. This implies that the second and the third order derivatives of ${\widetilde U}(yZ_T)$ are bounded, hence
the function ${\wt V}(y)=E{\wt U}(yZ_T)$ is three-times differentiable
and
$$
{\wt V}^{'''}(y)=E^Q{\wt U}^{'''}(yZ_T)Z_T^2.
$$
Since ${\wt V}(y)$ and $V(x)$ are conjugate, $V(x)$ is also three-times differentiable.

The duality relation (\ref{main}) takes is in this case the following form
\begin{equation}\label{dual2}
U^{'}(X_T(x))=yZ_T,\;\;\;\;\;X_T(x)=-{\wt U}^{'}(yZ_T), \;\;\;\;y=V^{'}(x).
\end{equation}
This relation implies that the function $X_T(x)$ is  two-times differentiable for all $\omega\in\Omega^{'}=(Z_T>0)$ with
$P(\Omega^{'})=1$ and differentiating the first equality in (\ref{dual2}) we have that
\begin{equation}\label{dif1}
U^{''}(X_T(x))X_T^{'}(x)=V^{''}(x)Z_T,
\end{equation}
\begin{equation}\label{dif2}
U^{'''}(X_T(x))(X_T^{'}(x))^2+U^{''}(X_T(x))X_T^{''}(x)=V^{'''}(x)Z_T.
\end{equation}
From (\ref{dual2}) and (\ref{dif1}) we obtain that
$$
X_T^{'}(x)=\frac{V^{''}(x)}{V^{'}(x)}\;\;\frac{U^{'}(X_T(x))}{U^{''}(X_T(x))}.
$$
By condition  $r1)$ and Proposition 1.2 from \cite{MT15}
$c_1\le -\frac{V^{''}(x)}{V^{'}(x)}\le c_2$. Therefore this implies that $X_T^{'}(x)$ is bounded, in particular
\begin{equation}\label{bou1}
\frac{c_1}{c_2}\le X_T^{'}(x)\le\frac{c_2}{c_1},
\end{equation}
where $c_1$ and $c_2$ are constants from (\ref{rav1}).

Comparing equations (\ref{dif1}) and (\ref{dif2}) we have that
\begin{equation}\label{comp}
X_T^{''}(x)+\frac{U^{'''}(X_T(x))}{U^{''}(X_T(x))}(X_T^{'}(x))^2=\frac{V^{'''}(x)}{V^{''}(x)}X^{'}_T(x).
\end{equation}
Since $E^QX_T^{'}(x)=1$ and $E^QX_T^{''}(x)=0$, taking expectations with respect to the measure $Q$ in equation (\ref{comp}) we get
\begin{equation}\label{comp2}
\frac{V^{'''}(x)}{V^{''}(x)}=E^Q\frac{U^{'''}(X_T(x))}{U^{''}(X_T(x))}(X_T^{'}(x))^2,
\end{equation}
which together with  (\ref{bou1}) and  condition $r1)$ implies that $\frac{V^{'''}(x)}{V^{''}(x)}$ is bounded.

Therefore, it follows from (\ref{comp}) that $X_T^{''}(x)$ is also bounded, hence $X_T'(x)$ is Lipschitz continuous.

Since the product of bounded Lipschitz continuous functions are Lipschitz continuous, it follows from (\ref{comp2}) that
$\frac{V^{'''}(x)}{V^{''}(x)}$ is Lipschitz continuous and (\ref{comp}) implies that $X^{''}_T(x)$ is also
Lipschitz continuous, since all terms in (\ref{comp}) are bounded and Lipschitz continuous.

{\bf Lemma 3.} Let the market be complete and  condition $r2)$ be satisfied.
Then the optimal wealth $X_T(x)$ is  three-times
differentiable, $X'_T(x)$ is strictly positive and  the derivatives $X_T^{'}(x), X_T^{''}(x)$
and $X'''_T(x)$ are uniformly bounded on every compact $[a,b]\in R$ .

{\it Proof.} Since $U(x)$ and $\wt U(y)$ are conjugate,  Condition $r2)$ implies that $\wt U(y)$ is also four times differentiable
and the derivatives of ${\widetilde U}(yZ_T)$ are bounded for any $y\in R$, hence
the function ${\wt V}(y)=E{\wt U}(yZ_T)$ is four-times differentiable.

Then $V(x)$ is also four-times differentiable, since $V'(x)$ is the inverse of $-\wt V'(y)$. Therefore,
the duality relation
$$X_T(x)=-\widetilde U'(V'(x)Z_T)$$
implies that the optimal wealth $X_T(x)$ is three-times
differentiable and  the derivatives $X_T^{'}(x), X_T^{''}(x)$ and $X'''_T(x)$ are bounded on every compact $[a,b]\in R$. Therefore
the derivatives $X_T^{'}(x), X_T^{''}(x)$ satisfy the local Lipschitz condition.

Besides,
$$X_T'(x)=-V''(x)Z_T\wt U''(V'(x)Z_T))>0$$ since $V''(x)<0$ and $\wt U''(y)>0.$

\be{cor}
The process $(X_t''(x),\;(t,x)\in [0,T]\times R)$ admits
 a continuous modification.
\ee{cor}
{\it Proof}. Since $X_t''(x)$ is a $Q-$martingale, by the Doob inequality and the mean value theorem we get
$$
E^Q\sup_{t\le T}|X_t''(x_1)-X_t''(x_2)|^2\le c_1E^Q|X_T''(x_1)-X_T''(x_2)|^2
$$
$$\le c_1|x_1-x_2|E^Q\sup_{\alpha\in[0,1]}|X_T'''(\alpha x_1+(1-\alpha)x_2)|^2\le
c_2|x_1-x_2|^2
$$
for some constants $c_1,c_2$. By the Kolmogorov theorem the map
$$R\ni x\to X_\cdot ''(x)\in C[0,T]$$
admits a continuous modification,
which implies the continuity of $X_t''(x)$ with respect to the variables $(t,x),\;P-$a.s.. \qed

{\bf Proposition 4.}  Assume that the market is complete and that  one of the
condition $r1)$ or $r2)$ is satisfied.

Then the optimal wealth $X_t(x)$, the optimal strategy $\pi_t(x)$ ($\mu^{\la S\ra}$-a.e.),
 martingale flows $\mathcal{M}(t,x)$ and  $\overline{\mathcal{M}}(t,x)$ are two-times continuously differentiable at $x$
for all $t$, $P-$a.s. and the coefficients of equation (\ref{sd}) satisfy the local Lipschitz condition.

{\it Proof}. Let first assume that condition
r1) is satisfied. According to Lemma 2 the optimal wealth $X_T(x)$ is two-times
differentiable and  the derivatives $X_T^{'}(x), X_T^{''}(x)$ are bounded and Lipschitz continuous.

To show an existence of $\pi'(x)$ we use the decomposition $X_T'(x)=1+\int_0^T\pi_r^{(x)}dS_r$ with some predictable $S$-integrable
integrand $\pi^{(1)}(x)$
and inequalities
\begin{eqnarray*}
  E^Q\int_0^T\left({\pi_t^{(1)}{(x+\vps)}-\pi_t^{(1)}{(x)}}\right)^2d\la S\ra_t&=&E^Q\left\la X'(x+\vps)-X'(x)\right\ra_T\\ =E^Q\left({X_T'(x+\vps)-X_T'(x)}\right)^2
  &\le&\vps^2 E^Q\max_{0\le s\le 1}|X''_T(x+s\vps)|^2\\
  \le\vps^2 Const,
\end{eqnarray*}
By the Kolmogorov theorem $\pi^{(1)}{(x)}$ is continuous with respect to $x$ $\mu^{\la S\ra}$-a.e.

Note that, if instead of r1) the condition r2) is satisfied, then we shall have that there exists a $\mu^{\la S\ra}$-a.e. continuous modification of $\pi^{(1)}(x)$ on each compact of $R$ which will imply an existence of continuous modification on the whole real line.

Thus by the stochastic Fubini Theorem
(see \cite{V} )
\begin{eqnarray*}
  x_2-x_1+\int_0^T(\pi_r(x_2)-\pi_r(x_1))dS_r &=& X_T(x_2)-X_T(x_1)\\
  =\int_{x_1}^{x_2}X_T'(x)dx&=& x_2-x_1+\int_0^T\int_{x_1}^{x_2}\pi_r^{(1)}{(x)}dxdS_r
\end{eqnarray*}
and consequently $\pi_r(x_2)-\pi_r(x_1)=\int_{x_1}^{x_2}\pi_r^{(1)}{(x)}dx$ $\mu^{\la S\ra}$-a.e..
 Hence $\pi^{(1)}{(x)}=\pi'(x)$ $\mu^{\la S\ra}$-a.e. and
 \begin{equation}\label{pi}
 X_T'(x)=1+\int_0^T\pi'_r(x)dS_r
 \end{equation}
 for all $x$ $P-$a.s..

 It follows from (\ref{pi}) and from the Fubini theorem that
 \begin{eqnarray*}
 X_t(x_2)-X_t(x_1)&=&x_2-x_1+\int_0^t(\pi_r(x_2)-\pi_r(x_1))dS_r \\
  =x_2-x_1+\int_0^t\int_{x_1}^{x_2}\pi_r'(x)dxdS_r&=&\int_{x_1}^{x_2}X_t'(x)dx
\end{eqnarray*}
for any $x_2\ge x_1$ $P-$a.s. and lemma $A3$ from \cite{MT15} implies that for each fixed $t$ there exists a modification of
$(X_t(x), x\in R)$ which is absolutely continuous with respect to the Lebesgue measure $dx$. Since $(X'_t(x), t\in [0,T])$ is a $Q$-martingale
\begin{equation}\label{lip}
|X'_t(x_2)-X'_t(x_1)|\le E^Q(|X'_T(x_2)-X'_T(x_1)|/F_t)\le C |x_2-x_1|
\end{equation}
for any $x_2\ge x_1$ $P-$a.s. and Lemma 2 and Corollary 1 imply that there exists
$\Omega'\subset\Omega,\;P(\Omega')=1$, such that  at each $\omega\in\Omega'$ the inequality (\ref{lip}) is fulfilled
for all $(t,x)$.

Since $EX''_T(x)=0$ and the market is complete we have
$X''_T(x)=\int_0^T\pi^{(2)}_r(x)dS_r$
for some predictable $S$-integrable integrand $\pi^{(2)}$. Similarly as above
one can show that $\pi^{(2)}(x)$ is continuous at $x$ $\mu^{<S>}$-a.e.,  $\pi^{(2)}(x)= \pi''(x)$ $\mu^{\la S\ra}$-a.e. and, hence $X''_t(x)$ admits the
 representation
 $$
X''_t(x)=\int_0^t\pi''_r(x)dS_r.
$$
Similarly we can show that one can choose a modification of $X_t(x)$ which is two-times differentiable and such that $X''(x)$ is Lipschitz continuous.

In case when instead of r1) the condition r2) is fulfilled $X''(x)$ will satisfy the local Lipschitz condition. So, in both cases (i.e., if condition r1) or r2) is satisfied) the coefficients of equation (\ref{sd}) will be locally Lipschitz continuous.

Since the market is complete $\overline{\cal M}(t,x)=V'(x)Z_t$ and it is evident that $\overline{\cal M}(t,x)$ is two-times continuously differentiable. Besides,
equality (\ref{calm}) implies that ${\cal M}(t,x)$ is also two-times continuously differentiable at $x$.

{\bf  Theorem 3.} Assume that the market is complete and that  one of the condition $r1)$ or $r2)$ is satisfied.
Then conditions a)-e) are fulfilled and the value function  $V(t,x)$ satisfies $BSPDE$ (\ref{ut6}).

{\it Proof.} It is evident that boundedness of $B_1(y)$ and $B_2(y)$ (defined by
(\ref{b1})) implies that the dual value function
$\wt V(t,y)=E(\wt U(y\frac{Z_T}{Z_t})/F_t)$ is two-times continuously differentiable. Since
$$
V''(t,x)=-\frac{1}{{\wt V}''(t,y)},\;\;\;y=V'(x),
$$
the value function $V(t,x)$ is also two-times continuously differentiable, hence condition a) is fulfilled.

It follows from Proposition 4 that under the presence assumptions all conditions of Propositions 2 and 3 are satisfied, therefore
these propositions imply that $V(t,x)$ satisfies conditions b) and c), hence $V(t,x)$ is a regular family of semimartingales.

Let us show that the condition e) is also satisfied.
By optimality principle (see \cite{MT10}) for any $t\in[0,T]$ the process
$(V(s,X_s(t,x)), s\ge t)$ is a martingale, where $X_s(t,x)=x+\int_t^s\pi_u(t,x)dS_u$ is the solution of the conditional optimization problem
(\ref{ut6}). This implies that $P$-a.s.
\begin{equation}\label{con1}
V(t,x)=E(V(s, X_s(t,x))/F_t).
\end{equation}
On the other hand using again the optimality principle we have
$$
V(t,X_t(x))=E(V(s,X_s(x))/F_t),
$$
 and substituting in this equality the inverse of the optimal capital $X_t(x)$ we get
\begin{equation}\label{con2}
V(t,x)=E(V(s,X_s(X_t^{-1}(x))/F_t).
\end{equation}
Since for any $t$ the function $(V(t,x), x\in R)$ is strictly convex, comparing (\ref{con1}) and (\ref{con2}) we obtain that $P$-a.s
 $X_{s}(t,x)=X_s(X_t^{-1}(x)).$  By continuity at $(t,x)$ of $X_t^{-1}(x)$ as a solution of SDE (\ref{sd}) we obtain that
 condition e) is satisfied.

Thus, all conditions of Theorem 3.1 from \cite{MT10} are satisfied which implies that $V(t,x)$ is a solution of the $BSPDE$
(\ref{ut6}). \qed

{\bf Corollary}. Let conditions of Theorem 3 be satisfied. Then the process
$$
\wt V(t,y)= E(\wt U(y\frac{Z_T}{Z_t})/F_t),\;\;\;t\in [0,T]
$$
satisfies the BSPDE (\ref{eqpded}).

{\it Proof.} According to
 Theorem 2 it is sufficient to verify that the process
$$
{\wt V}'(t,y)= E(\frac{Z_T}{Z_t}\wt U'(y\frac{Z_T}{Z_t})/F_t),\;\;\;t\in [0,T],
$$
 is a special semimartingale.

Let $\overline V(t.y)=E(Z_T\wt U'(y Z_T)/F_t)$. It is evident that ${\wt V}'(t,y)=\frac{1}{Z_t}\overline V(t, \frac{y}{Z_t}).$  But by the duality relation
(\ref{main})
 $\overline V(t.y)=E(Z_T\wt U'(y Z_T)/F_t)=-Z_tX_t(x)$ and the martingale field  $\overline V(t.y)$ is two-times differentiable by Proposition 4. Therefore
the It\^o-Ventzel formula implies that
$\frac{1}{Z_t}\overline V(t, \frac{y}{Z_t})$ is a special semimartingale, hence so is also the process  ${\wt V}'(t,y)$.

\

\end{document}

 where $b$ can be written similarly to
$-a$
\begin{multline*}
  b(r,x)
  =\overline{\cal M}'(r,\psi_r(x))\left(\frac{\pi'_r(\psi_r(x))\pi_r(\psi_r(x))}{X'_r(\psi_r(x))^2}-\frac{1}{2}\frac{X''_r(\psi_r(x))\pi_r^2(\psi_r(x))}{X'_r(\psi_r(x))^3}\right)
  \\+\frac{1}{2}\overline{\cal M}''(r,\psi_r(x))\frac{\pi_r^2(\psi_r(x))}{X'_r(\psi_r(x))^2}-\overline h_r(\psi_r(x))\frac{\pi_r(\psi_r(x))}
{X'_r(\psi_r(x))}.
\end{multline*}

The main example, where all conditions a)-e) are satisfied
is the case of exponential utility function
$$
U(x)=-e^{-\gamma x}
$$
with risk aversion parameter $\gamma\in(0,\infty)$. In this case $\widetilde{U}(y)=\frac{y}{\gamma}\big(\ln\frac{y}{\gamma}-1\big)$ and Assumption $d')$ is equivalent to the existence of $Q\in\mathcal{M}^e$ with finite relative entropy $EZ^Q_T\ln Z^Q_T$ (see e.g.  \cite{BF}).

In this case the corresponding  value function
is of the form $V(t,x)=-e^{-\gamma x}V_t$ and
\begin{equation}\label{utve}
V_t=\underset{\pi\in\Pi_x} {\text{\rm
essinf}}\text{ } E(e^{-\gamma(\int_t^T\pi_udS_u)}|{\mathcal
F}_t) \end{equation}
is a special semimartingale. Therefore, it is evident that conditions a)-e) are satisfied
and the BSPDE (\ref{ut6}) is transformed into a usual backward stochastic differential equation (BSDE). In particular, theorem 3.1 from \cite{MT10} implies
that $V(t,x)=-e^{-\gamma x}V_t$
where $V_t$ satisfies the BSDE
\begin{equation}\lbl{bexp}
V_t= V_0+\frac{1}{2}\int_0^t \frac{(\varphi_s+\lambda_sV_s)^2}
{V_s}d\langle M\rangle_s
+\int_0^t\varphi_sdM_s+L_t, \;\;\;V_T=1,
\end{equation}
where  $L$ is a local martingale strongly orthogonal to $M$ and the optimal wealth process is expressed as
\begin{equation}\label{frexp}
X_t(x)=x+\int_0^t\frac{\varphi_u+\lambda_uV_u} {\gamma V_u}dS_u.
\end{equation}

It was shown in \cite{S3} that under assumptions $1$ and an assumption of reasonable asymptotic elasticity (\ref{ae})
the optimal strategy $\pi(x)$ of the
problem (\ref{ut1}) in the class $\Pi_x$ exists.
It follows from \cite{S3} that Assumption $d')$ also implies the existence of optimal solution to the
conditional optimization problem (\ref{ut5}) (where the optimal strategy may depend on $t$ in general).